\documentclass[12pt]{article}
\usepackage{graphicx}
\newcommand{\beq}{\begin{equation}}
\newcommand{\eeq}{\end{equation}}
\newcommand{\bea}{\begin{eqnarray}}
\newcommand{\eea}{\end{eqnarray}}

\newcommand{\ha}{\hat{a}}

\renewcommand{\b}{\beta}
\renewcommand{\a}{\alpha}

\def\fun#1#2{\lower3.6pt\vbox{\baselineskip0pt\lineskip.9pt
  \ialign{$\mathsurround=0pt#1\hfil##\hfil$\crcr#2\crcr\sim\crcr}}}
\def\sq{{\vbox {\hrule height 0.6pt\hbox{\vrule width 0.6pt\hskip 3pt
   \vbox{\vskip 6pt}\hskip 3pt \vrule width 0.6pt}\hrule height 0.6pt}}}
\begin{document}
\begin{titlepage}
\begin{flushleft}
       \hfill                      {\tt hep-th/020****}\\
       \hfill                       FIT HE - 02-06 \\
\end{flushleft}

\renewcommand{\thefootnote}{\fnsymbol{footnote}}
\vspace*{3mm}
\begin{center}
{\bf\LARGE Scalar field localization on a brane with cosmological constant\\ }
\vspace*{5mm}

\bigskip

{\large Kazuo Ghoroku\footnote[2]{\tt gouroku@dontaku.fit.ac.jp}\\ }
\vspace*{2mm}
{
\large 
 Fukuoka Institute of Technology, Wajiro, Higashi-ku}\\
{
\large 
Fukuoka 811-0295, Japan\\}
\vspace*{5mm}

{\large Masanobu Yahiro \footnote[3]{\tt yahiro@sci.u-ryukyu.ac.jp} \\}
\vspace{2mm}
{
\large 
 Department of Physics and Earth Sciences, University of the Ryukyus,
Nishihara-chou, Okinawa 903-0213, Japan \\}

\vspace*{10mm}

\end{center}

\begin{abstract}
We address the localization of a scalar field, whose bulk-mass M is considered
in a wide range including the tachyonic region, 
on a three-brane. The brane with non-zero cosmological constant $\lambda$
is embedded in five dimensional bulk space. We
find in this case that the trapped scalar could have mass $m$
which has an upper bound and
expressed as $m^2=m_0^2+\alpha M^2\leq \beta |\lambda|$ with the calculable numbers
$m_0^2, \alpha, \beta$. We point out that this result would be
important to study the stability of the
brane and also the cosmological problems based on the brane-world.

\end{abstract}
\end{titlepage}

\section{Introduction}

The idea of brane-world ~\cite{RS1,RS2}
gives an alternative to the standard Kaluza-Klein (KK)
compactification via the localization of the 4d graviton on the brane
of Minkowski metric \cite{RS2} and also of de Sitter space
\cite{bre}.
It also opened a new way to the construction of
the hierarchy between four-dimensional Planck mass and
the electro-weak scale, and also for realization of the small
observable cosmological constant ($\lambda$)
with lesser fine-tuning \cite{ArHa,KaSc,GY}.

In order to develop this idea,
it would be necessary to study the localization of other fields which are the
ingredient of our world. There have been many approaches in the case of
$\lambda=0$, i.e. the Randall-Sundrum (RS) brane.
For finite $\lambda$, the analyses
have been concentrated on the massless mode, especially the graviton,
localization up to now.

Other than the graviton, there would be many kinds of fields
in the bulk since the five dimensional
bulk theory would be obtained by the reduction of the ten dimensional IIB
theory or eleven dimensional M theory. In any case, we would obtain
many number of scalar fields in the reduced five dimensional theory
\cite{PPN,BC}.
Many people have tried to understand the dynamical situation and
stable bulk configuration for a suitable brane world.
In these approaches, the scalar
fields play an important role, so it would be meaningful
to know their behaviour
in constructing the brane in the five
dimensional space. Further we notice in AdS$_5$ that negative mass
squared ($M^2$) is in general allowed. The value is bounded from below
$M^2\ge -4/L^2$, where $L$ is the radius of AdS space, for the case 
of $\lambda >0$  \cite{Bre}. The same
bound of the tachyon mass is also obtained from our analysis 
given here for non zero
$\lambda$.

From the above viewpoint, we have previously examined the localization
of the tachyonic scalar on the RS brane to discuss its stability \cite{GN1}.
Here we extend this analysis to the brane with finite $\lambda$.
In this case, the warp factor is deformed
compared to the case of RS brane due to the cosmological constant
and we find a mass gap for the KK mode.
Then it would be interesting to see how these points affect the localization
of the bulk massive fields. We can consider various other kinds
of bulk fields which have mass including the tachyonic field.

For negative $\lambda$,
it is known that the trapped graviton is massive on the brane
(AdS$_4$ brane) \cite{KR}, and it would be interesting
to see the situation for other trapped bulk field, especially for the
massive fields, on AdS$_4$ brane. From this viewpoint, we
study also the case of AdS$_4$ brane.

As a simple case to cover the problems stated above, we consider in this paper
the localization of massive scalar fields
by extending its bulk-mass to the tachyonic region. We could find a simple
mass relation between the mass for the trapped and the bulk-scalar, and
many implications would be obtained from it.

In Section 2, we give a brief review of brane solutions with non-zero
cosmological constant. They are obtained previously on the basis of
a simple ansatz imposed on the bulk metric \cite{bre}.
In Section 3, the localization of massive scalars on those
brane solutions are examined for both dS$_4$ and AdS$_5$ branes.
Concluding remarks and discussions are given in the final section.

\section{Brane solution with cosmological constan}

We start from the following five-dimensional gravitational action\footnote{
Here we take the following definition, $R_{\nu\lambda\sigma}^{\mu}
=\partial_{\lambda}\Gamma_{\nu\sigma}^{\mu}-\cdots$, 
$R_{\nu\sigma}=R_{\nu\mu\sigma}^{\mu}$ and $\eta_{AB}=$diag$(-1,1,1,1,1)$. 
Five dimensional suffices are denoted by capital Latin and the four
dimensional
one by the Greek ones.
},
\beq
    S = {1\over 2\kappa^2}\Bigg\{
      \int d^5X\sqrt{-G} (R -  2\Lambda + L_{m})
          +2\int d^4x\sqrt{-g}K\Bigg\}+S_{\rm b}, \label{action}
\eeq
where $L_{m}$ denotes the contribution from matter,
$K$  being the extrinsic curvature on the boundary.
The fields included in $L_{m}$ are not needed to construct the background 
of the brane. And the last term is the brane action,
\beq
    S_{\rm b} = -{\tau}\int d^4x\sqrt{-g}, \label{baction}
\eeq
Then the Einstein equation is solved in the following metric,
\beq
 ds^2= A^2(y)\left\{-dt^2+a^2(t)\gamma_{ij}(x^i)dx^{i}dx^{j}\right\}
           +dy^2  \, \label{metrica},
\eeq
where the coordinates parallel to the brane are denoted by $x^{\mu}=(t,x^i)$,
 $y$ being the coordinate transverse to the brane. The position of the brane
is taken at $y=0$. We restrict our interest here to the case of a
Friedmann-Robertson-Walker type
(FRW) universe. Then, the three-dimensional metric $\gamma_{ij}$
is described in Cartesian coordinates as
\beq
  \gamma_{ij}=(1+k\delta_{mn}x^mx^n/4)^{-2}\delta_{ij},  \label{3metric}  
\eeq
where the parameter values $k=0, 1, -1$ correspond to a
 flat, closed, or open universe respectively.

\vspace{.3cm}
When  considering the metric (\ref{metrica}), we obtain 
the following reduced equations \cite{bre}:
\beq
  ({\dot{a_0}\over a_0})^2+{k\over a_0^2}=A'^2+{\Lambda\over 6}A^2
          =D,  \label{Einstein2}
\eeq
where $D$ is a constant being independent of $t$ and $y$. 
In view of the boundary condition at the brane position,
\beq
  {A'(0+)-A'(0-)}=-{\kappa^2\tau\over 3}A(0), \label{bound2}
\eeq
one gets
\beq
   D\equiv \lambda = \kappa^4\tau^2/36+\Lambda/6 , \label{4cos}
\eeq 
if  $A(0)=1$. The normalization condition
$A(0)=1$ does not affect the generality of our discussion.

\vspace{.3cm}
When $\lambda >0$, we obtain the solution for 
$a_0(t)$ which represents inflation of three space. Here we give the
simple one, which is used hereafter, for the case of $k=0$,
\beq
 a_0(t)=e^{H_0t},
\eeq
where the Hubble constant is represented as $H_0=\sqrt{\lambda}$.
%When $\lambda >0$, we obtain the solution for 
%$a_0(t)$ as given in the Table 1, where the Hubble constant is represented as
%$H_0=\sqrt{\lambda}$ and $\alpha_i$ are constants. These solutions 
%represent inflation of three space.

%\vspace{.3cm}
%\begin{center}
%\begin{tabular}{|c|c|c|c|}%\hline\hline
%\multicolumn{4}{c}{Table 1.} \\ \hline
%{}& $k=0$ &$k=1$ &$k=-1$ \\ \hline
%$a_0(t)$ & $e^{H_0t}$ & ${1\over H_0}\cosh(H_0t+\alpha_1)$ &
%    ${1\over H_0}\sinh(H_0t+\alpha_2)$ \\ \hline
%\end{tabular}
%\end{center}
%\vspace{.3cm}

As for $A(y)$, for any value of $k=0, \pm1$,
we obtain the same solution. We notice that
$a_0(t)$ has nothing to do with the problem of
localization and it depends only on the form of $A(y)$, as we will see
below. For $\Lambda <0$, $A(y)$ is solved as
\beq
  A(y)= {\sqrt{\lambda}\over\mu} \sinh[\mu(y_H-|y|)] \, \label{metrica4}
\eeq
\beq
  \mu=\sqrt{-\Lambda/6}, \qquad \sinh(\mu y_H)=\mu/\sqrt{\lambda}.
               \label{constads}
\eeq
where $y_H$ represents the position of the horizon in the bulk.
This solution represents a brane at $y=0$. The configuration is taken
to be $Z_2$ symmetric with respect to the reflection, $y\to -y$. 
%%%%%%%%%%%%%%%%%%%%%%%

\vspace{.5cm}

When  $\Lambda$ is positive, the solution for  
 $a_0(t)$ is the same as above, but  
 $A(y)$ becomes different.
One has
\beq
 A(y)={\sqrt{\lambda}\over \mu_d}\sin[\mu_d(y_H-|y|)],
          \label{desit}
\eeq 
\beq
  \mu_d=\sqrt{\Lambda/6}, \qquad \sin(\mu_d y_H)=\mu_d/\sqrt{\lambda}.
               \label{const1}
\eeq
Here $y_H$ represents the position of the horizon in the bulk dS$_5$, where
there is no spatial boundary as in AdS$_5$.
This configuration represents a brane with dS$_4$ embedded in the
bulk dS$_5$ at $y=0$. The $Z_2$ symmetry is also imposed. 

\vspace{.5cm}

As for $\lambda<0$, we obtain a real solution 
\beq
 a_0(t)={1\over H} \sin(H t), \quad H=\sqrt{-\lambda}
 \label{a0-ads4}
\eeq
for $k=-1$, but imaginary ones for others. The scale factor $a_0$ should be 
positive, and only the real solution can satisfy this condition, 
when $0<t<\pi/H$. Thus, the real solution works only for the finite period.
If all the evolution of our universe is governed by the solution, 
the period $\pi/H$ should be larger than the present cosmic age. 
In this case, $H$ is smaller $10^{-3}$ eV. 
The AdS$_4$ brane is obviously prohibited for $\Lambda>0$, because of 
(\ref{4cos}). For $\Lambda <0$, $A(y)$ is obtained as
\beq
  A(y)= {H\over\mu} \cosh[\mu(y_{H}-|y|)] \, \label{metrica-AdS4}
\eeq
\beq
  \mu=\sqrt{-\Lambda/6}, \qquad \cosh(\mu y_{H})=\mu/H .
               \label{const-AdS4}
\eeq
Note that the bulk has no horizon in this case. 
The configuration is $Z_2$ symmetric again.

%\newpage
\section{Localization of massive scalars}

The case of $\lambda=0$ is studied widely, 
and it is known that there is no localization of massive-mode
except for the resonances which would decay into the bulk.
While there is a possibility of trapping the massive mode of
$m^2<9\lambda/4$ for the case of $\lambda>0$.
So we discuss here the case of $\lambda>0$, where we consider both solutions
for $\Lambda>0$ and $\Lambda<0$. 

For the sake of the simplicity, consider a scalar, $\Phi$, in the bulk
with mass $M$. Its equation of motion is represented by
\beq
   ({\sq}_5-M^2)\Phi=0  \,, \label{mscalar}
\eeq
where $\sq_5$ denotes the bulk laplacian. For the background metric of $k=0$,
\beq
  ds^2= A(y)^2(-dt^2+a_0(t)^2\delta_{ij}dx^idx^j)+dy^2, \label{metrica6}
\eeq
this operator can be decomposed as
\beq
 \sq_5\equiv {1\over \sqrt{-G}}\partial_N \sqrt{-G} G^{NL}\partial_L
       = {1\over A^2(y)}\sq_4 + 
        (\partial_y^2+{4\over A}\partial_yA\partial_y) \,, \label{mlapl}
\eeq
and $\sq_4$ denotes the laplacian on the (1+3) dimensional brane.

In this case, Eq.~(\ref{mscalar}) is written by expanding
$\Phi$ in terms of the four-dimensional continuous mass eigenstates:
\beq
 \Phi=\int dm \varphi_m(t,x^i)\phi(m,y) \, , \label{eigenex}
\eeq
where the mass $m$ is defined by
\beq
  -\sq_4\varphi=\ddot{\varphi}_m+3{\dot{a}_0\over a_0}\dot{\varphi}_m
           +{-\partial_i^2\over a_0^2}\varphi_m=-m^2\varphi_m , \label{masseig}
\eeq
and $\dot{}=d{}/dt$. 
When $a_0=1$, we get the usual relation, 
$-k^2=-\eta^{\mu\nu}k_{\mu}k_{\nu}=m^2$, by taking as
$\varphi_m=e^{i k_{\mu}x^{\mu}}$,  $m$ representing the four-dimensional mass.
The explicit form of the solution of Eq.~(\ref{masseig}) is given as
\beq
 \varphi_m(t)=e^{-3H_0t/2}\left\{ c_1J_{\nu}(q)+c_2J_{-\nu}(q)\right\},
\label{phi-m}
\eeq
\beq
  \nu=\sqrt{{9\over 4}-m^2/H_0^2}\, ,\qquad q=e^{-Ht}\sqrt{{k^2\over H_0^2}},
\eeq
where $H_0=\sqrt{\lambda}$ as given in the previous section. However we do not
need to use this exact form hereafter. Here we need only the explicit form
of $\phi(m,y)$, and its equation is obtained as 
\beq
  {\phi}''+4{A'\over A}{\phi}'
           +{m^2\over A^2}\phi=M^2\phi , \label{warp}
\eeq
where ${}'=d{}/dy$.

\vspace{.5cm}
The localization is studied by solving Eq.(\ref{warp}). It is
rewritten into the form of one-dimensional 
Schr\"{o}dinger-like equation with the eigenvalue $m^2$,
\beq
 [-\partial_z^2+V(z)]u(z)=m^2 u(z) , \ \label{warp3}
\eeq
where the potential $V(z)$ is given as
\beq
 V(z)={9\over 4}(A')^2+{3\over 2}AA''+A^2M^2,
\eeq 
and we introduced $u(z)$ and $z$ defined as 
$\phi=A^{-3/2}u(z)$ and $\partial z/\partial y=\pm A^{-1}$.
The potential $V(z)$ in this equation is 
determined by $A(y)$, and it contains a $\delta$-function attractive
force at the brane position to trap the lowest mode of the fields.

Before solving Eq.~(\ref{warp3}), we comment on some point on 
the lowest eigenvalue of $m^2$ by giving it as follows,
\beq
  m^2=\int dz u^{*}(z)[-\partial_z^2+V(z)]u(z)\equiv
       <u|[-\partial_z^2+V(z)]|u>,  \label{masseig3}
\eeq
where $u(z)$ denotes the normalized eigenfunction, $<u|u>=1$.
Using the estimations given before in the cases of $\lambda=0$ and 
$M^2\neq 0$ \cite{GN1,DRT},
$\lambda\neq 0$ and $M^2=0$ \cite{bre,KR,Mie}, we can expand $m^2$ near
$M^2=0$ as
\beq
    m^2=m_0^2+\alpha M^2.   \label{massrel}
\eeq
Here $m_0^2=0$ for $\lambda\geq 0$ \cite{GN1,bre} and 
$m_0^2=\beta |\lambda|$ for
$\lambda<0$, where $\beta$ is a calculable positive number \cite{KR,Mie}.
The coefficient $\alpha$ is positive since $dm^2/dM^2=<u|A^2|u>$ 
is positive at any $M^2$, and it can be estimated as
\beq
  \alpha=<u|A^2|u>|_{M^2=0}~~ >0.
\eeq

\vspace{.5cm}
From these comments, we can imagine the $M^2$-dependence of $m^2$ which
should be given as the solution of (\ref{warp3}).
In the following subsections, we can assure these points by examining 
the spectrum of $u(z)$ for both positive and negative $\lambda$.
We also investigate how massive particles are localized on the brane. One more
point to be noticed is that $m^2$ of the bound state must
be less than the minimum of the potential $V(z)$ in the sense of one 
dimensional quantum mechanics. Then there is an upper bound for the bound
state mass.

\vspace{.5cm}
\subsection{de Sitter Brane}

For $\lambda>0$, 
there is a possibility of trapping the massive mode of
$m^2<9\lambda/4$, since $V(z)$ tends to the value at $z=\infty$.
The positive $\lambda$ is allowed for both $\Lambda>0$ and $\Lambda<0$. 

\vspace{.5cm}
For this case, the potentials are given by using the solutions
given in the previous section as follows,

{\bf For $\Lambda >0$ :}
$z=\rm{sgn}(y)(\lambda)^{-1/2}\ln(\cot[\mu_d(y_H-|y|)/2])$ and
\beq
 V(z)= {15\over 4}\lambda [{a_1-1\over\rm{cosh}^2(\sqrt{\lambda}z)}+{3\over 5}]
      -{\kappa^2\tau\over 2}\delta(|z|-z_0) , \label{pot2}
\label{V-dS5}
\eeq
\beq
  a_1={4M^2\over 15\mu_d^2}={8M^2\over 5\Lambda} \, , \qquad
 z_0={1\over\sqrt{\lambda}}\rm{arccosh}({\sqrt{\lambda}\over \mu_d}).
\eeq

{\bf For $\Lambda <0$ :} In this case,
$z=\rm{sgn}(y)(\lambda)^{-1/2}\ln(\rm{coth}[\mu(y_H-|y|)/2])$ and
\beq
 V(z)= {15\over 4}\lambda [{1+a_2\over\rm{sinh}^2(\sqrt{\lambda}z)}+{3\over 5}]
      -{\kappa^2\tau\over 2}\delta(|z|-z_0) . \label{pot3}
\label{V-AdS5}
\eeq
\beq
 a_2=4M^2/(15\mu^2), \qquad
 z_0={1\over\sqrt{\lambda}}\rm{arcsinh}({\sqrt{\lambda}\over \mu}).
\eeq
%%%%%%%%%
Here $z=z_0$ corresponds to $y=0$, the position of the brane. (Note that 
$\sqrt{\lambda}/\mu_d \ge 1$, because of Eq.~(5).)
We notice several points with respect to these potentials. 
%%%%%%%%%%%%%%%%%%%%%%%%%%%%
%\begin{enumerate}
%\item 

\vspace{.5cm}
(i) We should take as $\tau >0$ to realize attractive 
$\delta$-function force at the brane since this is a necessary condition
for the graviton localization.
(ii) The analytic part of the potential depends on the value
of $M$, and it determines the value of $V(z_0)$ as,
\beq
  V(z_0)=V(a_i=0,z_0)+M^2 \, , \qquad 
      V(a_i=0,z_0)={9\over 4}\lambda-{5\over 8}\Lambda .
\eeq
When $V(a_i=0,z_0)<0$, the graviton could not be trapped on the brane \cite{bre},
then we consider the case of $V(a_i=0,z_0)>0$.
(iii)
Typical shapes of $V(z)$ are shown in the Fig.1. The shape of (A) is therefore
abandoned here. So, we consider the forms of (B) and (C) hereafter.
It should be noticed that the type (C)
is similar to the volcano-shape as seen in the case of $\Lambda<0$ and
$\lambda=0$, RS solution.

In any case, the potentials monotonically approach to
the asymptotic value
\beq
      V_{\infty}={9\over 4}\lambda
\eeq
at $z=\infty$ or at the horizon $y=y_H$. 
Then the mass confined on the brane would be expected in the region
\beq
     m^2 < V(z_0).
\eeq
And continuum KK modes appear for $m^2>V_\infty$.
%%%%%%%%%%%%%%% Fig %%%%%%%%%%%%%%
\begin{figure}[htbp]
\begin{center}
\voffset=15cm
  \includegraphics[width=8cm,height=7cm]{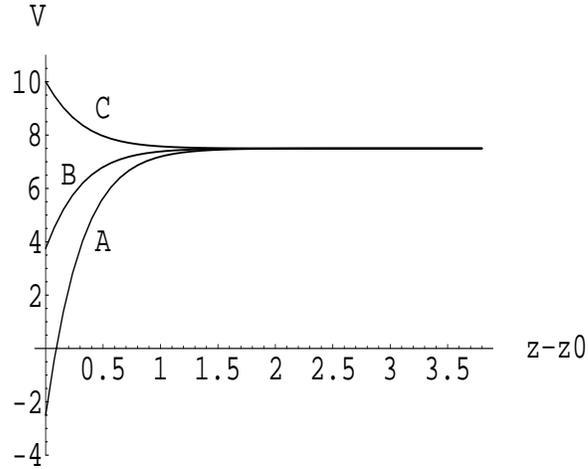}
\caption{The curves A, B and C show the three typical 
finite part of $V(z)$ for the case of $\lambda>0$.
The scale is arbitrary. The left hand end-points
of each curve represents $V(z_0)$, where $\delta$-function attractive 
potential appears, but it is suppressed here.}
\end{center}
\end{figure}
%%%%%%%%%%%%%%% Fig %%%%%%%%%%%%%%

\vspace{.5cm}
The bound states for the case of potentials
given above are examined in terms of the explicit
solution of Eq.~(\ref{warp3}).

Firstly, it is given for $\Lambda>0$ as
\beq
 u(z)=c_1X^{-id} {}_2F_1(b_1,b_2;c;X)+c_2X^{id}{}_2F_1(b_1',b_2';c';X), 
        \label{solds}
\eeq
where $c_{1,2}$ are  constants of integration and
\beq
 X={1\over\rm{cosh}^2(\sqrt{\lambda}z)}, \quad 
      d={\sqrt{-9+4m^2/\lambda}\over 4},   \label{para1}
\eeq
\beq
  b_1={1\over 4}-\sqrt{1-{15\over 16}a_1}-id, 
   \quad b_2={1\over 4}+\sqrt{1-{15\over 16}a_1}-id, 
   \quad c=1-2id, \label{para2}
\eeq
\beq
  b_1'={1\over 4}-\sqrt{1-{15\over 16}a_1}+id, 
     \quad b_2'={1\over 4}+\sqrt{1-{15\over 16}a_1}+id, 
     \quad c'=1+2id. \label{para3}
\eeq
Here ${}_2F_1(b_1,b_2;c;X)$ denotes the Gauss's hypergeometric function.
It follows from this solution that $u(z)$ oscillates 
with $z$ when $m>3\sqrt{\lambda}/2$,
where the continuum KK modes appear. 
The coefficients $b_i$ and $b_i'$ are complex for $a < 16/15$, but this
complexity is not essential for the oscillatory behaviour of $u(z)$.
While for $m<3\sqrt{\lambda}/2$,
$u(z)$ should decrease rapidly for large $z$ since the mode given 
in this region
should be a bound state. Then one must take $c_2=0$.

\vspace{.3cm}
Nextly, the solution for $\Lambda<0$ is obtained as
\beq
 u(z)=\bar{c}_1Y^{-id} {}_2F_1(\bar{b}_1,\bar{b}_2;c;-Y)
      +\bar{c}_2Y^{id}{}_2F_1(\bar{b}_1',\bar{b}_2';c';-Y), 
     \label{solds2}
\eeq
where $\bar{c}_{1,2}$ are integration constants and
\beq
 Y={1\over\rm{sinh}^2(\sqrt{\lambda}z)}.
\eeq
Here $(\bar{b}_1,\bar{b}_2)$ and $(\bar{b}_1',\bar{b}_2')$ 
are obtained from $(b_1,b_2)$ and $(b_1',b_2')$ given in (\ref{para2})
and (\ref{para3}) by replacing $a_1\to -a_2$. Other parameters, $c, c'$ and $d$
are the same with those given in (\ref{para1}) $\sim$ (\ref{para3}).
Here we comment on the lower bound of mass-aquared of the tachyon which is
given by the energy positivity in AdS \cite{Bre}. 
The same bound is seen from
our solution (\ref{solds2}), where two terms are complex conjugate each other
when the mass is within the mass bound, $M^2\geq -4\mu^2$. However this relation is broken
for $M^2<-4\mu^2$, then we can not find any well-defined energy operator
for this solution (\ref{solds2}) out of the mass bound. 
This mass bound is in general given as $M^2<-(d-1)^2\mu^2/4$ for AdS$_d$.
Similar analysis
would be performed for dS$_5$, but energy is not well-defined in this 
space-time. So we don't extend this discussion for dS case.

The bound state is expected in the region, $m<3\sqrt{\lambda}/2$, where
$u(z)$ must decrease rapidly at large $z$. Then we take $\bar{c}_2=0$ 
as above case.

%%%%%%%%%%%%%%%%%
\vspace{.3cm}
In any case, these solutions must
satisfy the boundary condition at $z=z_0$,
\beq
 u'(z_0)=-{\kappa^2\tau\over 4}u(z_0),  \label{boundzero}
\eeq
because of the existence of $\delta$-function in the potential.
And the eigenvalue $m^2$ of the bound state is given as the solution
of this equation (\ref{boundzero}).

The results are shown in the Fig.2. As expected,
we find the solution of positive $m^2$ in the region,
$m^2<V_{\infty}={9\over 4}\lambda$, for $M^2>0$. Furthermore, the 
tachyonic bound states ($m^2<0$) are found for negative $M^2$.
%%%%%%%%%%%%%%% Fig %%%%%%%%%%%%%%
\begin{figure}[htbp]
\begin{center}
\voffset=15cm
   \includegraphics[width=8cm,height=7cm]{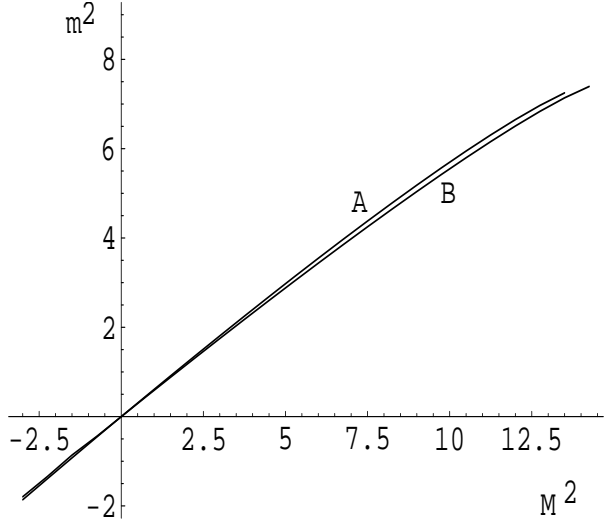}
 \caption{
The solutions $m^2$ of $u(z_0)+4u'(z_0)/(\kappa^2\tau)=0$ at chosen values of 
$M^2$ are shown for ($\Lambda>0$, $\mu_d=1$)
and ($\Lambda<0$, $\mu=1$) by the curves A and B respectively.
Here we take $\lambda=10/3$.}
\end{center}
\end{figure}
%%%%%%%%%%%%%%% Fig %%%%%%%%%%%%%%

\vspace{.5cm}
For the bound state given above, we should show the normalizability 
for the localization in the following sense
\beq
 \int_0^{y_H}dy A^2(y)\Phi^2(m,y)= 
     \int_{z_0}^{\infty}dz~ u^2(m,z)< \infty,   \label{norm}
\eeq
for the bounded-mode solution $u(m,z)$ given above. It is easy to see
that this condition is satisfied if $m^2<9\lambda /4$, then the solutions
are normalizable.

\vspace{.3cm}
The results for the case of $\lambda>0$ are summarized
as follows; For both cases of $\Lambda>0$ and $\Lambda<0$, (i)
we can find the bound state in the expected region, 
$0\leq m<3\sqrt{\lambda}/2$, for positive $M^2$.
(ii) In the case of $M^2<0$, the tachyonic bound state of $m^2<0$ appears
as presented as in the case of $\lambda=0$ \cite{GN1}. This point will
be important in considering the stability-problem of brane-world.
 
%%%%%%%%%%%%%%%%%%%%%%%%%%%%%%%%%%
\vspace{.5cm}

\vspace{.5cm}

%%%%%%%%%%%%%%%%%%%%%%%%%%%%%%%%%% AdS brane %%%%%%%%%%
\subsection{Anti de Sitter Brane}

In this subsection, we consider the case of AdS brane ($\lambda<0$). 
Such brane is obviously allowed only for AdS bulk ($\Lambda < 0$), 
because of (\ref{4cos}).
For the Randall-Sundrum brane ($\lambda =0$), it is known that 
tachyons can live in AdS$_5$ bulk with keeping the bulk stable 
\cite{Bre}.
This property persists also for the AdS brane, as shown later.
Thus, three types of particles, massive and massless particles and 
tachyons, can reside in AdS$_5$ bulk with AdS$_4$ brane. 
We then investigate their localization systematically.

Using $\partial z/\partial y= A^{-1}$ for $A(y)$ of 
Eq. (\ref{metrica-AdS4}), we obtain a new coordinate $z$ as 
$ z= {\rm sgn}(y) \arctan{ \{ \sinh{ \mu(|y|-y_H) } \} }/H $.
Discussions made below are parallel for both cases of positive and 
negative $y$. We then consider the case of positive $y$ only. 
As for positive $y$, the potential $V$ in Eq. (\ref{warp3}) is 
\beq
 V(z)= {15\over 4} H^2 [\frac{ \ha }{ \cos{^2(H z)} }-{3\over 5}]
      -{\kappa^2\tau\over 2}\delta(z-z_0) , \label{pot-ads}
\eeq
\beq
  \ha=1+a \, ,\qquad
  a={4M^2\over 15\mu^2}=-{ 8M^2\over 5\Lambda} \, , \qquad
  H=\sqrt{-\lambda} \, , \qquad
  z_0=- {1\over H} \arccos ({H \over \mu}) . 
\eeq
Here $z$ varies in a range $z_0 \le z \le \pi/2H$, and 
$z_0$ is a value of $z$ at $y=0$, indicating 
the brane resides in $z=z_0$.
Note that $ 0 \le \b \equiv H/\mu \le 1$ because of Eq.~(\ref{4cos}).
The potential contains a $\delta$-function force with the same form as 
in the case of dS brane ($\lambda>0$). Hence, the solution of Eq. (\ref{warp3}) satisfies 
the same boundary condition at $z=z_0$ as Eq.(\ref{boundzero}).
Again, an attractive $\delta$-function force ($\tau >0$) guarantees that 
particles are trapped on the brane.

%%%%%%%%%%%%%%% Fig %%%%%%%%%%%%%%
\begin{figure}[htbp]
\begin{center}
\voffset=15cm
   \includegraphics[width=8cm,height=7cm]{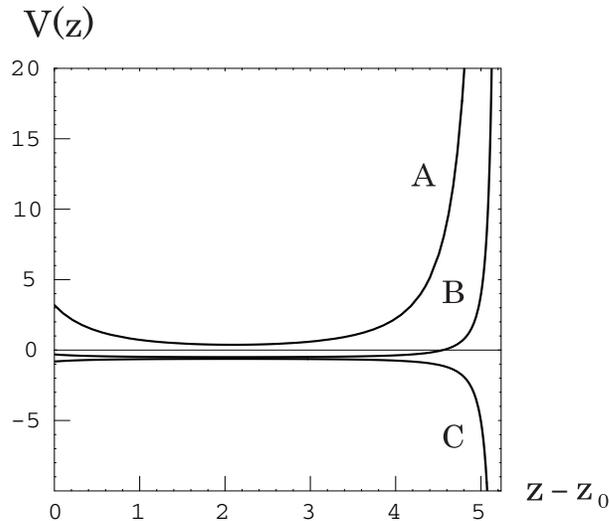}
 \caption{
The $z$ dependence of $V(z)$ in a case of 
$H=0.5$ and $\mu=1$. Here $z$ varies from $z_0=-2.09$ to 
$z_1=\pi$. 
Three lines A, B, C show the result for  $\ha=1, 1/15, -1/15$
($M^2=0, -7/2, -4$), respectively. } 
\end{center}
\end{figure}
%%%%%%%%%%%%%%% Fig %%%%%%%%%%%%%%

The analytic part of $V(z)$ depends on $M$ through $\ha$.
Three typical cases are shown in Fig. 3.
Types (A) and (B) show the part for $\ha=1$ and 1/15, 
respectively. As for $\ha \ge 0$, thus, 
the analytic part is divergent at $z=z_1$, 
where $z_1=\pi/2H$. All the eigenstates $u(z)$ of Eq.(\ref{warp3}) are 
then bound states satisfying 
\beq
u(z_1)=0.
\label{bc1}
\eeq
On the other hand, Type (C) is the potential for $\ha=-1/15$ 
corresponding to $M^2=-4$.
For $\ha < 0$, the potential is strongly attractive near $z=z_1$. 
This means that in general the resultant bound state is localized 
not only near $z=z_0$ but also near $z=z_1$. 
Thus, particles can not be trapped only at the brane in this case.
We then consider only the case of $\ha \ge 0$,
corresponding to $M^2 \ge -15\mu^2/4$.

The analytic part of $V$ has a minimum at $z=0$. It behaves as 
$ V(0)=3H^2/2+H^2M^2/\mu^2$. 
When the eigenvalue $m^2$ of Eq.(\ref{warp3}) is smaller than the minimum 
$V(0)$, the corresponding bound state is localized on the brane.  
When $m^2 \ge V(0)$, on the other hand, the bound state spreads in the bulk.
The mass confined on the brane is then expected in the region
\beq
     m^2 < V(0).
\label{trap}
\eeq

%%%%%%%%%%%%%%% Fig %%%%%%%%%%%%%%
\begin{figure}[htbp]
\begin{center}
\voffset=15cm
   \includegraphics[width=8cm,height=7cm]{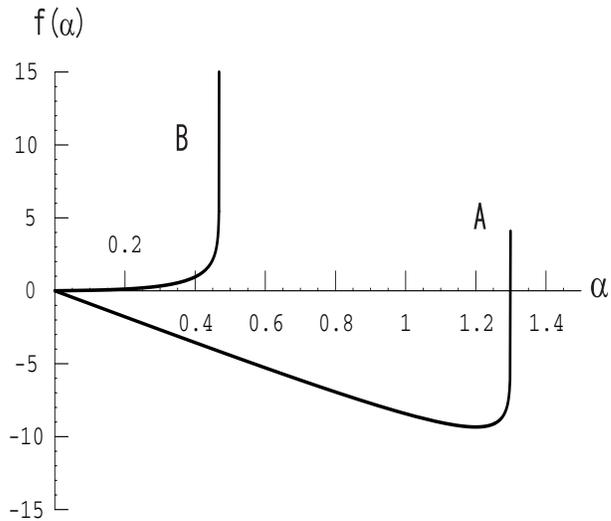}
 \caption{
The $\a$ dependence of $f$. 
Two lines A, B show the result for  $\b=0.5$ and 0.95, respectively.
 } 
\end{center}
\end{figure} 
%%%%%%%%%%%%%%% Fig %%%%%%%%%%%%%%

First, we consider the limit of small $\ha$ and analyze analytically  
the localization of particles there. This analysis gives a good
insight to the localization mechanism for finite $\ha$, as shown 
later. In the limit, only tachyon
can live in the bulk, because $M^2 < 0$ there. 
Comparing two curves (A) and (B) in Fig. 3, we find that 
the analytic part of $V$ tends to a well potential: 
$V(z)=-9H^2/4$ for $z_0 < z <z_1$ 
and $V(z)=\infty$ for $z=z_1$. The solution of Eq. (\ref{warp3}) then takes
the form $u(z) =c_1 \exp(-\a \mu z)+c_2 \exp(\a \mu z)$, 
where $\a^2=-(m/\mu)^2-9\b^2/4 $ and $\b=H/\mu$. 
The relation between $\a^2$ and $m^2$ indicates that 
$\a$ is real for the solution to satisfy 
the trapping condition (\ref{trap}). 
Imposing two boundary conditions, Eqs. (\ref{boundzero}) 
and (\ref{bc1}), on the solution, 
we can determine eigenvalues from Eq. (\ref{warp3}). 
The lowest eigenvalue $m^2$, relevant to the localization, satisfies 
\beq
 f(\a) \equiv \log[{ {3 \over 2} \sqrt{1-\b^2} + \a  \over 
{3 \over 2} \sqrt{1-\b^2} - \a }] 
 -2 \a [ {\pi \over 2\b} +{\arccos{(\b)} \over \b} ] =0 .
\label{eigenvalue}
\eeq
The equation has a solution $\a=0$, but it leads to a trivial solution 
$u(z)=0$. 
We then define the left hand side of Eq.(\ref{eigenvalue}) as $f(\a)$
and investigate the $\a$ dependence 
in order to find out a nontrivial solution. 
Figure 4 shows the $\a$ dependence for two typical cases of $\b=0.5$ and 0.95.
It is found from the figure that there exists a nontrivial solution for 
$\b=0.5$, but not for $\b=0.95$. This is understood, as follows. 
For any $\b$, $f(\a)$ is zero at $\a=0$ and infinity 
at $ \a = 3\sqrt{1-\b^2}/2$, and the second derivative of $f(\a)$ 
with respect to $\a$ is positive in the range $0 < \a < 3\sqrt{1-\b^2}/2$. 
So, Eq.(\ref{eigenvalue}) has a nontrivial solution 
when $df(\a)/d\a <0 $ at $\a=0$, 
but no solution 
when $df(\a)/d\a \ge 0 $ at $\a=0$.
The first derivative of $f$  has a simple form 
$ 4/(3\sqrt{1-\b^2})-2\{\pi/(2\b)+\arccos{(\b)}/\b \}$ at $\a=0$,
and it is negative at $0<\b<0.9440$ and positive at $0.9440<\b<1$.
Hence, there is a solution satisfying the trapping condition (\ref{trap})
for $0<\b<0.9440$, 
but not for $0.9440<\b<1$.
The solution thus found has a negative $m^2$ 
at $M^2=-15\mu^2/4$. 
In the limit of small $\ha$, therefore, 
the tachyon living in the bulk is trapped 
on the brane as a tachyon, only when $0<\b<0.9440$.

Next, we analytically investigate the $M^2$ dependence of the 
lowest eigenvalue $m^2$. 
The lowest eigenvalue and the corresponding normalized eigenstate, $u$, 
satisfy Eq. (\ref{masseig}). 
Differentiating both the sides of the equation with respect to $M^2$ leads to 
\beq
   {d m^2 \over d M^2}=<u|{d V(z) \over d M^2}|u> ,
\label{m-dep}
\eeq
since $<du/dM^2|u>+<u|du/dM^2>=0$ because of  $<u|u>=1$. 
The right hand side of Eq. (\ref{m-dep}) is easily calculated into 
$<u(M,z)|{\rm sec}^2(Hz)|u(M,z)> H^2/\mu^2$, and found to be larger than 
$H^2/\mu^2$ because of $\cos{(Hz)} \le 1$ at any $z$ between $z_0$ and $z_1$. 
This leads to an relation 
\beq
{d m^2 \over d M^2} > {d V(0) \over d M^2}= {H^2 \over \mu^2}.
\label{m-V}
\eeq
As for $0.9440<\b<1$, the lowest eigenvalue $m^2$ is above $V(0)$ 
at the limit of small $\ha$, as mentioned above.
The relation (\ref{m-V}) guarantees that 
$m^2 > V(0)$ at any positive $\ha$, 
or at any $M^2$ larger than the lower bound $-15\mu^2/4$, 
indicating no localization of particles on the brane 
at any $M^2$. For $0<\b<0.9440$, on the other hand, $m^2$ is below 
$V(0)$ in the limit of small $\ha$, as mentioned above. 
As $M^2$ increases from the lower bound,
$m^2$ overtakes $V(0)$ because of the relation (\ref{m-V}). Thus, 
particles are trapped on the brane at least for $M^2$ 
near the lower bound.

For the case of positive $\ha$, the Schr\"{o}dinger-like equation 
(\ref{warp3}) has a general solution 
$u(z)=\bar{c}_1F(z)+\bar{c}_2G(z)$ for 
\beq
 F(z)=W^{-id} {}_2F_1(\bar{b}_1,\bar{b}_2;c;W) , \quad
 G(z)=W^{id}{}_2F_1(\bar{b}_1',\bar{b}_2';c';W), 
     \label{sol-ads}
\eeq
where $\bar{c}_{1,2}$ are integration constants and
\beq
 W={1 \over {\rm cos}^2(Hz)}, \quad 
d={\sqrt{-9-4m^2/H^2}\over 4}. 
\eeq
Here $(\bar{b}_1,\bar{b}_2,c)$ and $(\bar{b}_1',\bar{b}_2',c')$ 
are obtainable from $(b_1,b_2,c)$ and $(b_1',b_2',c')$, 
defined in (\ref{para2}) and (\ref{para3}), by replacing $a_1 \to -a$. 
After the replacement, Eqs. (\ref{para2}) and (\ref{para3}) 
show that 
$G(z)$ and $F(z)$ are complex conjugate to each other for $M^2 \ge -4 \mu^2$, 
but not for $M^2 < -4 \mu^2$. 
As stated above, it indicates that 
the ground state of AdS$-5$ bulk is stable only at $M^2 \ge -4 \mu^2$. 
A similar discussion is also possible for AdS brane. 
The corresponding stability condition for the brane is $m^2 < 9H^4/4$.

The eigenmode of Eq.(\ref{warp3}) has to satisfy 
the two boundary conditions, (\ref{boundzero}) and (\ref{bc1}):
\beq
{ G'(z_0)+{3 \over 2} \sqrt{\mu^2-H^2}G(z_0)  \over 
  F'(z_0)+{3 \over 2} \sqrt{\mu^2-H^2}F(z_0) } =
\lim_{z \to z_1} { G(z) \over F(z) } .
\label{bc-ads4}
\eeq
Both $F(z)$ and $G(z)$ are divergent at $z=z_1$, although the ratio is 
finite.  As a result of the dangerous behavior 
of $F(z)$ and $G(z)$ near $z=z_1$, it is hard to obtain any eigenmode 
numerically with Eq. (\ref{bc-ads4}). So we solve 
the Schr\"{o}dinger-like equation (\ref{warp3}) numerically 
with the Runge-Kutta method which propagates solutions satisfying the 
initial condition (\ref{bc1}) from $z_1$ to $z_0$. Among numerical 
solutions, each with different $m^2$, eigenmodes are chosen so that 
the solution can satisfy the boundary condition (\ref{boundzero}) 
at $z=z_0$.

%%%%%%%%%%%%%%% Fig %%%%%%%%%%%%%%
\begin{figure}[htbp]
\begin{center}
\voffset=15cm
   \includegraphics[width=8cm,height=7cm]{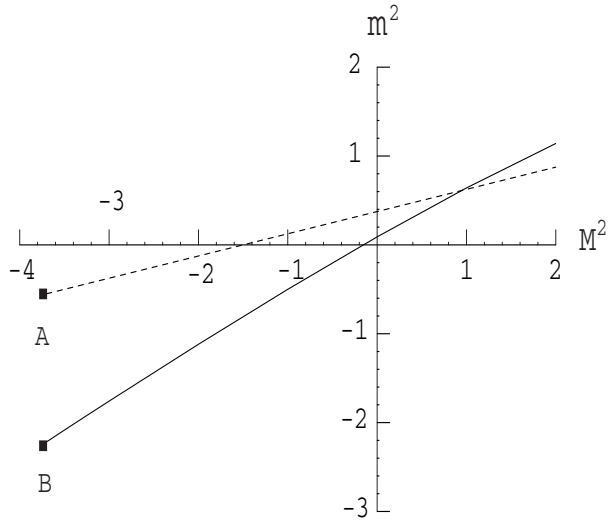}
 \caption{
The lowest eigenvalue $m^2$ and the potential minimum $V(0)$ 
as a function of $M^2$. The former is shown by a solid line and the latter 
is by a dashed one. Points A and B are $V(0)$ and $m^2$ in the limit of 
$\ha \to 0$. Parameters taken are $\mu=1$ and $\b=H/\mu=0.5$. } 
\end{center}
\end{figure}
%%%%%%%%%%%%%%% Fig %%%%%%%%%%%%%%

Figure 5 shows the resulting lowest eigenvalue $m^2$ as a function of $M^2$. 
Here we take $\mu=1$ and $\b=H/\mu=0.5$, as an example of $\b$ belonging 
to the region $0<\b<0.9440$. 
Points A and B stand for $V(0)$ and $m^2$, respectively, at 
$M^2=-3.75$ corresponding to the limit of $\ha \to 0$.
As expected, $m^2$ is below $V(0)$ there. 
As $M^2$ is increased
from the lower bound, $m^2$ catches up with $V(0)$ at $M^2=0.95$. 
Thus, tachyons and massless and massive particles are confined in the brane 
when $-3.75<M^2<0.95$. At $-0.15<M^2< 0$, especially, a tachyon is trapped 
on the brane as either a massless or massive particle. 
The resultant $M^2$ dependence of $m^2$ is a simple form 
$m^2=c_1+c_2 M^2$, where $c_i$ are positive. 
A graviton with $M^2=0$ is confined as a massive particle with 
small $m^2=c_1$. This is consistent with the previous analyses on graviton 
\cite{KR}. As for $0.9440<\b<1$, on the other hand, 
numerical calculations 
shows no localization of particles at any $M^2$, as expected.

%%%%%%%%%%%%%%%%%%%%%%%%%%%%%%%%%%

%%%%%%%%%%%%%%%%  Conclusion %%%%%%%%%%%%%%%%%%
\section{Concluding remarks}

We have examined 
the localization of massive scalar on both dS and AdS
branes, where mass $M$ of scalar particle is varied from 
the massless- and massive-particle region $M^2 \ge 0$ 
to the tachyonic region $M^2 < 0$. 
As a bulk space, we consider both cases, AdS$_5$ and dS$_5$. 
The dS$_5$ bulk, with positive 5d cosmological constant ($\Lambda >0$), has 
recently attracted interest in connection with the
proposed dS/QFT correspondence \cite{Nojiri}. The AdS$_5$ bulk 
is not only interested in the context of AdS/CFT correspondence, 
but also important in the sense that the bulk 
is derivable from superstring through the dimensional reduction.
As a characteristic of AdS$_5$ bulk, it is found that 
the ground state of the bulk is stable only when $M^2 \ge - 4|\Lambda|/6$, 
independently of brane, for both dS and AdS branes. 
Thus, even tachyons can reside in AdS bulk, with keeping the vacuum stable, 
when $M^2 \ge - 4|\Lambda|/6$.

As for dS brane, with positive 4d cosmological constant ($ \lambda>0 $), 
the continuous KK modes of scalar fluctuation are found in 
the mass range $m^2 > 9 \lambda/4$, 
for both AdS$_5$ and dS$_5$ bulks. 
The mass gap for the KK modes is 
a characteristic of dS brane.

For such dS brane, massive scalar is localized on the brane 
as a massive mode. A mass $m$ of the mode is related to $M$ 
as $m^2=c M^2$ for a calculable positive constant $c$, 
when $m^2 \ll |\Lambda|$. 
The localized mode has to be found 
in the mass gap, so massive scalar can be localized 
on dS brane only when $M^2 < 9 \lambda/4c$. 
In the case of the Randall-Sundrum brane with $\lambda=0$, 
on the other hand, 
there is no mass gap for the continuous KK modes, so that 
any massive scalar can not be localized on the brane.

As for AdS brane, which is allowed only for AdS$_5$ bulk, 
massive scalar with $M^2>0$ 
is also localized on the brane as a massive mode, 
and the localized mode is related 
to $M^2$ as $m^2=c_1 M^2 + c_2$ for positive constants $c_i$. 
In the limit of small $M^2$, the formula 
shows that graviton is trapped on AdS brane 
as a massive particle \cite{KR}.

Extending the analyses mentioned above to negative $M^2$,
we can find tachyonic ($m^2<0$) localized modes, 
for both cases of dS and AdS branes.
The relation of the mode to $M^2$ is the same as 
in the case of massive mode: $m^2=c M^2$ for dS brane and 
$m^2=c_1 M^2 + c_2$ for AdS brane. Furthermore, it is found that 
the ground state of brane is stable at $m^2>0$ for the case of dS brane 
and at $m^2>-9|\lambda|/4$ for the case of AdS brane.

As for dS brane embedded in AdS$_5$ bulk, 
the formula on $m^2$ shows that once tachyons exist in the bulk, 
the tachyonic mode also appears on the brane 
and makes the brane unstable. In this sense, tachyons can not exist 
in AdS$_5$ bulk. 
For AdS brane embedded in AdS$_5$ bulk, on the other hand,
the formula on $m^2$ shows a possibility that tachyons are trapped on the brane as either a massless or massive mode, but graviton is not 
localized on the AdS brane as a massless mode. 
Therefore, we can conclude that tachyons should be 
prohibited in AdS$_5$ bulk for both cases of AdS and dS branes.
It is quite an interesting issue why and how tachyons are prohibited 
when the five-dimensional theory is reduced from superstring.

As a natural statement, we can say that the observed acceleration of 
the present universe is induced by small positive $\lambda$.
At the same time,  it is expected from the recent
observations that some amount of cold dark matter coexists with 
the cosmological constant. The coexistence would be explainable
from the brane-world viewpoint. 
If a small but finite $\lambda$ exists, massive scalar living in the bulk 
can be trapped on the present universe as a massive mode, 
when $m < 3\sqrt{\lambda}/2 \sim 10^{-3} $ eV. 
The trapped particles are so called `dark'
in the sense that they interact with ordinary matters on the
brane only through gravitation. 
As a result of the weak interaction, it is likely that the trapped particles 
are also `cold' in the sense that they are hardly thermalized.
Thus, this scalar could be 
a candidate for the cold dark matter. 
As an important fact, it should be stressed that the coexistence occurs only
when the positive cosmological constant exists.

\end{document}